*Article*

# The Impact of National Culture on Innovation: A Comparative Analysis between Developed and Developing Nations during the Pre- and Post-Crisis Period 2007–2021

Han-Sol Lee [1,*], Sergey U. Chernikov [1], Szabolcs Nagy [2] 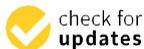 and Ekaterina A. Degtereva [1]

[1] Department of Marketing, Peoples' Friendship University of Russia, Miklukho-Maklaya, 6, 117198 Moscow, Russia
[2] Marketing and Tourism Institute, University of Miskolc, H-3515 Miskolc-Egyetemváros, Hungary
* Correspondence: li-kh@rudn.ru

**Abstract:** This empirical study investigates the impact of the Hofstede cultural dimensions (HCD) on the Global Innovation Index (GII) scores in four different years (2007, 2009, 2019 and 2021) to compare the impacts during the pre- and post-crisis (financial and COVID-19) period by employing ordinary least square (OLS) and robust least square (Robust) analyses. The purpose of this study is to identify the impact of cultural factors on the innovation development for different income groups during the pre- and post-crisis period. We found that, in general, the same cultural properties were required for countries to enhance innovation inputs and outputs regardless of pre- and post-crisis periods and time variances. The significant cultural factors (driving forces) of the innovation performance do not change over time. However, our empirical results revealed that not the crisis itself but the income group (either developed or developing) is the factor that influences the relationship between cultural properties and innovation. It is also worth noting that cultural properties have lost much of their impact on innovation, particularly in developing countries, during recent periods. It is highly likely that in terms of innovation, no cultural development or change can significantly impact the innovation output of developing countries without the construction of the appropriate systems.

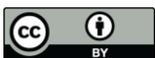

## 1. Introduction

Innovation plays an important role in promoting economic progress and competitiveness in both developed and developing countries (Şener and Sarıdoğan 2011). Many governments place innovation at the heart of their growth strategies (Patanakul and Pinto 2014). Nowadays innovation encompasses social, business and technical innovations (Dawson and Daniel 2010). Innovation in emerging economies is crucial to inspire people, which is especially true for the next generation of entrepreneurs and innovators (Reddy 2011).

Culture is a vital basis for innovation (Kaasa and Vadi 2008) and has a significant positive impact on it (Lažnjak 2011; Rinne et al. 2012; Taylor and Wilson 2012; Andrijauskiene and Dumciuviene 2017; Prim et al. 2017; Cox and Khan 2017; Yun et al. 2017; Handoyo 2018; Bukowski and Rudnicki 2019; Tekic and Tekic 2021; Espig et al. 2021). There are several studies exploring how culture affects innovation (Sun 2009); however, the evolution of cultural dimensions that influence countries' innovation performance over time and the impact of the crisis on them have not yet been studied.

Theoretically, in general, some specific cultural features are continuously manifested as desirable for innovation development. However, during a crisis, society should be tightly cooperating under strong leadership to efficiently respond to sudden changes and be quickly normalized from it. Thus, we identify whether different cultural properties are required for innovation development during pre- and post-crisis periods. In addition, the impact of cultural factors can be variant depending on the income level of a country, which is highly correlated with the level of innovation systems. The aim of this paper is to explore the relationships between Hofstede's cultural dimensions and the innovation performance of developing and developed countries before, during and after crisis as represented by the outbreak of the COVID-19 pandemic. In this study, we also address the existing research gap by responding to the research question concerning the evolution of the cultural dimensions affecting innovation performance over time. In terms of methodology, this study adopted multiple regression to measure the impact of cultural factors on innovation development based on mathematical accuracy. In particular, alongside ordinary least square (OLS) estimation, we further used robust estimation (the least median of squares method) to better deal with outliers.

The remaining parts of the paper are composed as follows. Section 2 is dedicated to a literature review on Hofstede's 6D model and the Global Innovation Index (GII) and hypothesis development. Section 3 describes data and methodology. Section 4 presents results and discussion. Section 5 includes conclusions and policy implications.

## 2. Literature Review

We intend to explore the links between national culture and innovation by examining the relationship between Hofstede's 6D model of national culture and the Global Innovation Index. The relationship between innovation and the specific dimensions of the Hofstede model—based on the summary of the findings of previous researches—are also discussed in this section.

### 2.1. Hofstede's 6D Model

Geert Hofstede's 6-D model of national culture is one of the most popular and widely accepted valid constructs to measure culture (Kirkman et al. 2006). It contains six dimensions, each of them assessed on a scale from 0 to 100: power distance, individualism versus collectivism, masculinity versus femininity, uncertainty avoidance, long-term orientation (previously Confucian dynamism) and indulgence versus restraint (Hofstede et al. 2010; Hofstede 1980).

Individualism (IDV) is the extent to which people feel independent, autonomous personalities as opposed to interdependent members of a tightly-knit society (collectivism). Individualism, which means that individual decisions are preferred, is not synonymous with egoism. In individualist societies, autonomy and personal initiatives are more frequent than in collectivist cultures in which people know their socially determined place and are loyal to a larger group (Hofstede 2022).

Power distance (PDI) versus closeness refers to the extent to which people accept or reject hierarchies and the unequal distribution of power in a society (Beugelsdijk and Welzel 2018). Low levels of power distance are manifested by strong centralization of power, decision-making and control, rigid hierarchies, lack of trust and less communication, which is often only one-way (Hofstede 2022).

Masculinity (MAS) refers to success, competition, achievement and the importance of material rewards versus cooperation, quality of life and caring. In a masculine society, which is primarily quantity-oriented, career, financial rewards and success are the top priorities. In a more tolerant, quality-focused feminine society, trust, caring for others, solidarity, cooperation and sympathy are significantly more important than competition (Hofstede 2022).

Uncertainty avoidance (UAI) versus acceptance refers to what extent people are averse to uncertainty and ambiguity. Uncertainty avoidance is very much related to anxiety and distrust of the unknown. Avoidance refers to how much people prefer existence under well-organized and highly predictable circumstances. However, in cultures accepting uncertainty, people are able to handle unplanned situations and are usually not afraid of failure. As a result, there are many novel ideas in uncertainty-accepting cultures (Hofstede 2022).

Long-term orientation (LTO) is associated with change. The long-time-oriented culture is future-oriented and pragmatist, while the short-time-oriented culture is past and present oriented, and honoring traditions and maintaining social norms prevails (Hofstede 2022). It must be noted that Minkov et al. (2018) criticized this dimension and proposed the use of "flexibility versus monumentalism" instead, which is a theoretically more focused and coherent dimension of national culture.

The sixth dimension is indulgence (IVR), which refers to expressing emotions and enjoying the good things in life and is the opposite of restraint (Hofstede 2022). In an indulgent culture, people are impulsive, socializing, having fun and enjoying life, whereas in a restrained culture, people often feel that life is hard, suppress emotional impulses and have a need for discipline and strict codes of conduct (Beugelsdijk and Welzel 2018).

## 2.2. The Global Innovation Index

The Global Innovation Index (GII) is a metric that has been assessing the performance of the innovation ecosystem of economies around the globe since 2007. GII, which aims to provide comprehensive picture of innovation, includes 81 indicators that measure the political environment, education, infrastructure and knowledge creation in each country (WIPO 2021). With its rich data metrics at the index, sub-index or indicator level, GII is widely used to monitor innovation performance and benchmark developments against countries within the same region or income group classification.

The GII—adapting the broad definition of innovation elaborated in the Oslo Manual (OECD—Organisation for Economic Co-Operation and Development 2018)—was developed to identify indicators and methodologies that can better capture the complexity and richness of innovation in society and go beyond traditional methods of measuring innovation, which typically only include indicators such as the number of research articles and the amount of money spent on research and development (R&D). It is an important tool for policy makers as its very detailed database provides an abundance of indicators for fine-tuning innovation policy.

As far as the GII conceptual framework is concerned, the overall GII ranking is based on two equally important sub-indices: the Innovation Input Sub-Index (IISI) and the Innovation Output Sub-Index (IOSI). IISI consists of five pillars that support and facilitate innovative activities in the economy: institutions, human capital and research, infrastructure, market sophistication and business sophistication. The first pillar, institutions, looks at the institutional framework of the economy including the political, regulatory and business environment. The next pillar, human capital and research, deals with education, tertiary education and research and development (R&D). The third pillar, infrastructure, refers to information and communication technologies (ICTs), the general infrastructure and ecological sustainability. The market sophistication pillar investigates credit, investment, trade, diversification and market scale. Business sophistication, the last pillar of innovation input, also includes three sub-pillars: knowledge workers, innovation linkages and knowledge absorption. IOSI is made up of two pillars representing the result of innovative activities in a society: knowledge and technology outputs and creative outputs. Knowledge creation, knowledge impact and knowledge diffusion are the three sub-pillars of knowledge and technology outputs, while creative outputs include intangible assets, creative goods and services as well as online creativity (WIPO 2021). Sub-pillars are computed using the weighted average of each indicator and are normalized on a scale from 0 to 100. The pillar scores are calculated using the weighted average of the sub-pillar scores. IISI and IOSI are equally important in calculating the overall GII scores, and therefore the same weight is assigned to them. GII economy rankings is based on the overall GII score, which is the average of IISI and IOSI (WIPO 2021).

Despite criticisms of GII, such as the overemphasis on factors that are not integral to innovation (Dašić et al. 2020), it is still one of the most appropriate and complex measures of a country's innovation performance.

*2.3. The Relationship between Culture and Innovation*

There is an abundance of research using Hofstede's cultural dimensions to better understand the role of culture in the success of innovation. Table 1 summarizes the different methodologies and databases that were used in publications focusing on investigating the relationship between innovation and Hofstede's cultural dimensions.

**Table 1.** Methodologies and databases used in research on the relationship between Hofstede's dimensions and innovation.

| Source | Methodology | Database Used |
|---|---|---|
| Kaasa and Vadi (2008) | correlation analysis | Patenting intensity–Eurostat's Regio database and European Social Survey 2007, 20 countries |
| Sun (2009) | meta-analysis | Innovation capability acc. to Porter and Stern (2001) and Hofstede's 4D (Hofstede et al. 1991) |
| Rinne et al. (2012) | multivariate multiple linear regression | Global Innovation Index (GII) |
| Halkos and Tzeremes (2013) | data envelopment analysis (DEA) | European Innovation Scoreboard 2007 (25 EU countries) |
| Prim et al. (2017) | multiple linear technical regression | Global Innovation Index (GII) 2016, 72 countries |
| Andrijauskiene and Dumciuviene (2017) | correlation analysis and multiple regression analysis | European Innovation Scoreboard 2016, 27 EU countries |
| Handoyo (2018) | bivariate correlation analysis and multiple regression analysis | Hofstede's national culture Index and Global Competitiveness Index, 77 countries |
| Jourdan and Smith (2021) | multiple regression analysis, factor analysis | the Global Innovation Index (GII) (143 countries), the Global Entrepreneurship Index (GEI = 119), the Global Creativity Index (GCI = 118), and Bloomberg 50 most innovative countries (B50) |
| Tekic and Tekic (2021) | fuzzy-set qualitative comparative analysis (fsQCA) | Global Innovation Index 2019, 91 countries |
| Espig et al. (2021) | multiple linear regression | Hofstede's national culture database, the Global Innovation Index and population data from the World Bank database from 2015 to 2018, 71 countries. |

Higher growth-rate and stronger tendency for innovation are associated with low uncertainty avoidance and low power distance (Hofstede 1980). Kaasa and Vadi (2008) investigated the relationships between Hofstede's cultural dimensions and the innovation potential of measured by the number of patent applications. They found power distance, uncertainty avoidance, family-related collectivism and masculinity negatively correlated with innovation measured by patenting intensity. After a meta-analysis of the previous studies, Sun (2009) found that individualism, power distance and uncertainty avoidance are correlated with national innovation capability. Lažnjak (2011) investigated the dimensions of national innovation culture in Croatia and found that power distance, collectivism and uncertainty avoidance are associated with lower innovation capacity. Rinne et al. (2012) found negative relationship between PDI and GII innovation scores. However, they detected strong positive relationship between individualism and GII innovation scores and no relationship between GII and uncertainty avoidance. Halkos and Tzeremes (2013) found higher power distance and uncertainty avoidance negatively affect the countries' innovation efficiency. Prim et al. (2017) analyzed the relationship between Hofstede's cultural dimensions and the Global Innovation Index 2016 using multiple linear technical regressions. They found individualism, long-term orientation and individualism are all positively associated with the capacity for innovation of the countries. Open innovation can motivate individualism by increasing individual creativity. Individualism, in turn, motivates open innovation through individual creativity. In contrast, collectivism reduces the complexity motivated by open innovation (Yun et al. 2017). Bukowski and Rudnicki (2019) revealed that individualism, long-term orientation and flexibility have strong positive impacts on national innovation intensity. They also point out that there is no consistent pattern regarding the impact of culture on national innovation rates, which should be taken into account when looking for effective innovation strategies and policies. Handoyo (2018) examined the relationship between national culture and the capacity of the nation to innovate and found that individualism, long-term orientation and indulgence were positively and significantly associated with national innovative capacity. Power distance and uncertainty avoidance negatively influence national innovative capacity. Moreover, masculinity had no relationship with national innovative capacity. Jourdan and Smith (2021) investigated how Hofstede's national culture dimensions influence indices of nations' creativity, entrepreneurship and innovation. They reduced Hofstede's six cultural dimensions to three major factors: heteronomy–autonomy, gratification, and competition–altruism. Using multiple regression analysis, they found that heteronomy–autonomy and gratification are predictors of nations' innovation. Tekic and Tekic (2021), employing the lens of neo-configurational theory, combined Hofstede's dimensions into five cultural profiles and investigated how multiple Hofstede's dimensions interact and combine to influence national innovation performance using the fuzzy-set qualitative comparative analysis (fsQCA). They found that cultures with high power distance, collectivism and short-term orientation are associated with low national innovation performance. However, individualism and long-term orientation are correlated with higher capacity for innovation.

Table 2 shows the empirical evidence of the relationships between Hofstede's cultural dimensions and the innovation performance of countries.

**Table 2.** The relationships between Hofstede's cultural dimensions and the innovation performance of countries.

| Hofstede's Cultural Dimension | Relationship with Innovation | | |
|---|---|---|---|
| | **Negative** | **No Relationship** | **Positive** |
| **Power Distance (PDI)** | Shane (1993); Hofstede (1980); Kaasa and Vadi (2008); Sun (2009); Lažnjak (2011); Rinne et al. (2012), Halkos and Tzeremes (2013); Prim et al. (2017); Andrijauskiene and Dumciuviene (2017); Handoyo (2018); Tekic and Tekic (2021); Espig et al. (2021) | | |
| **Individualism (IDV)** | | | Shane (1993); Sun (2009); Lažnjak (2011); Rinne et al. (2012); Taylor and Wilson (2012); Andrijauskiene and Dumciuviene (2017); Prim et al. (2017); Cox and Khan (2017); Yun et al. (2017); Handoyo (2018) Bukowski and Rudnicki (2019); Tekic and Tekic (2021); Espig et al. (2021) |
| **Masculinity (MAS)** | Kaasa and Vadi (2008); Espig et al. (2021) | Handoyo (2018) | |
| **Uncertainty Avoidance (UAI)** | Hofstede (1980); Shane (1993) Kaasa and Vadi (2008); Sun (2009); Lažnjak (2011); Halkos and Tzeremes (2013) Prim et al. (2017); Andrijauskiene and Dumciuviene (2017); Handoyo (2018); Espig et al. (2021) | Rinne et al. (2012) | |

**Table 2.** *Cont.*

| Hofstede's Cultural Dimension | Relationship with Innovation | | |
|---|---|---|---|
| | Negative | No Relationship | Positive |
| Long-term Orientation (LTO) | | | Cox and Khan (2017); Prim et al. (2017); Handoyo (2018); Bukowski and Rudnicki (2019); Tekic and Tekic (2021); Espig et al. (2021) |
| Indulgence (IDV) | | | Prim et al. (2017); Andrijauskiene and Dumciuviene (2017); Handoyo (2018); Espig et al. (2021) |

### 2.4. Hypothesis Development

Taylor and Wilson (2012) demonstrated that although individualism generally fosters innovation development, a certain type of collectivism spurs innovation activities. In detail, according to the research, a type of collectivism, such as patriotism and nationalism, helps to promote innovation at a national level. In particular, in the post-crisis period, we assume that the strong social ties among citizens could critically act to recover the societies back to the pre-crisis level or even promote them forward and even overtake countries that were previously better off.

**Hypothesis 1 (H1).** *Individualism is positively associated with innovation in the pre-crisis period (2007 and 2019) but negatively associated with innovation in the post-crisis period (2009 and 2021).*

Our study further investigates the relationship of leadership (by using PDI) and the innovation development based on the crisis management. During the crisis, the normal rhythm of life is disturbed, and unpredictable strikes require the leaders to make emergent and immediate decisions to reduce social ambiguities (Danielsson 2013). A society based on a strong leader and high-power distance suppresses the creativity of people as providing a small room for voices during the normal period. However, a charismatic leadership could be efficient and relatively easily calm the confusions of societies during the crisis period, the time when the expectations from the leader is higher than the non-crisis period (Bligh et al. 2004; Mumford et al. 2007), giving firm and fast directions that all members should follow.

**Hypothesis 2 (H2).** *High power distance negatively affects innovation in the pre-crisis period (2007 and 2019) but positively affects innovation in the post-crisis period (2009 and 2021).*

Another issue is the speed of development. Generally, it is visible that both innovation development and culture diffusion have significantly increased over the last 40 years. The cross-border successes of innovation incubators like Silicon Valley and Shenzhen are widely known and have been a copy model for boosting research across the world. At the same time, the cross-culture impact and enrichment is greatly stimulated through internet access to patterns of different cultures, predominantly through entertainment products. While innovation development has a lot of impacting factors (Derindag et al. 2021), it is obvious that culture is a more complex and inflexible system. The comparably faster pace of innovation development patterns raises the question of whether their correlation with cultural traits remains stable throughout the time or shows a change.

**Hypothesis 3 (H3).** *The significant cultural factors (driving forces) of innovation performance change over time.*

On the other hand, this study further explores a level of impacts of cultural factors on the innovation development between different income groups. The important feature of modern innovations is that they are predominantly costly, and not only in financial terms. While developed states have the necessary infrastructure in terms of university networks, utilities, security, legal systems, transport, research facilities, venture capital and such, the developing states may lack a number of important components there (Hu et al. 2021). Therefore, even having a very innovative-friendly culture, developing states may have lower ranking in GII due to sheer institutional or material underdevelopment.

**Hypothesis 4 (H4).** *The impact of cultural factors on innovation is stronger in developed countries than in developing countries.*

### 3. Data and Methodology

For the data and regression analysis, we used EViews ver. 12 as software. In our study, the GII data for the year 2007, 2009, 2019 and 2021 are selected for 77 countries (See Table A2). GII datasets in the year 2007 and 2009 are scaled 1–7, while those of the year 2019 and 2021 are scaled 0–100 (which has been used since 2011), as GII has expanded and diversified its indicators and developed calculation methodologies to better treat missing values, outliers, normalization, several years of references and so forth. On the other hand, HCD are time-invariant variables. The descriptive data are presented in Table 3. To ascertain the use of ordinary least square (OLS) as a regression methodology, we clarified the normal distribution of datasets based on skewness, kurtosis and Jarque–Bera (J-B) normality test. In particular, to make datasets in a normal distribution, we took a natural logarithm in GII datasets for the years 2007 and 2009. Although all the variables are within the acceptable range of a normal distribution in terms of skewness (between -2 and +2) and kurtosis (between -7 and +7) (Byrne 2010; Hair et al. 2010), IDV did reject the null hypothesis of J-B normality test, indicating that IDV is non-normally distributed at a 5% significance level. Excluding IDV, as the *p*-value of J-B normality test is above 0.5, all the other dependent and independent variables are normally distributed at a 5% significance level.

**Table 3.** Descriptive data.

| Variables | N | Mean | St. Dev. | Min. | Max. | Skewness | Kurtosis | J-B Prob. |
|---|---|---|---|---|---|---|---|---|
| Ln(GII_07) | 77 | 1.048066 | 0.275793 | 0.506818 | 1.757858 | 0.190045 | 2.411639 | 0.455179 |
| Ln(GII_09) | 77 | 1.225659 | 0.208799 | 0.862890 | 1.581038 | 0.270223 | 1.819480 | 0.066926 |
| GII_19 | 77 | 41.49000 | 11.83708 | 22.87000 | 67.24000 | 0.251603 | 1.972037 | 0.122293 |
| GII_21 | 77 | 39.99481 | 12.23134 | 19.70000 | 65.50000 | 0.199386 | 1.994216 | 0.152915 |
| PWI | 77 | 62.31169 | 20.77211 | 11.00000 | 100.0000 | −0.347020 | 2.402276 | 0.260321 |
| IDV | 77 | 42.49351 | 23.14919 | 10.00000 | 91.00000 | 0.465942 | 1.900330 | 0.035688 |
| MAS | 77 | 48.19481 | 18.96676 | 5.000000 | 100.0000 | 0.025022 | 3.441407 | 0.728641 |
| UAI | 77 | 67.58442 | 21.95890 | 8.000000 | 100.0000 | −0.487146 | 2.296691 | 0.098644 |
| LTO | 77 | 47.72727 | 23.02152 | 7.000000 | 100.0000 | 0.193820 | 2.034557 | 0.176178 |
| IVR | 77 | 44.70130 | 21.85422 | 0.000000 | 97.00000 | 0.176450 | 2.248526 | 0.330988 |

Source: composed by authors.

To describe the Pearson's correlation coefficients illustrated in Table 4, excluding MAS, all the other variables of the HCD present correlations with the GII variables with statistical significance. At a 1% significance level, PWI has a strong negative correlation with the GII variables, while IDV has a strong positive correlation with the GII variables. UAI shows a slight adverse-movement with the GII variables at a 1% significance level, while LTO and IVR show a slight co-movement with the GII variables at a 1% significance level.

**Table 4.** Pearson's correlation matrix.

|  | Ln(GII_07) | Ln(GII_09) | GII_19 | GII_21 | PWI | IDV | MAS | UAI | LTO | IVR |
|---|---|---|---|---|---|---|---|---|---|---|
| Ln(GII_07) | 1.000000 —- | | | | | | | | | |
| Ln(GII_09) | 0.898987 (0.0000) | 1.000000 —- | | | | | | | | |
| GII_19 | 0.886044 (0.0000) | 0.943466 (0.0000) | 1.000000 —- | | | | | | | |
| GII_21 | 0.887254 (0.0000) | 0.934071 (0.0000) | 0.994022 (0.0000) | 1.000000 —- | | | | | | |
| PWI | −0.607805 (0.0000) | −0.669686 (0.0000) | −0.636594 (0.0000) | −0.616166 (0.0000) | 1.000000 —- | | | | | |
| IDV | 0.695116 (0.0000) | 0.708043 (0.0000) | 0.698211 (0.0000) | 0.672882 (0.0000) | −0.723676 (0.0000) | 1.000000 —- | | | | |
| MAS | 0.070911 (0.5400) | −0.067462 (0.5599) | −0.033487 (0.7725) | −0.037622 (0.7453) | 0.141916 (0.2183) | 0.053301 (0.6452) | 1.000000 —- | | | |
| UAI | −0.372284 (0.0009) | −0.352717 (0.0017) | −0.337539 (0.0027) | −0.312942 (0.0056) | 0.259129 (0.0229) | −0.234234 (0.0403) | 0.007274 (0.9499) | 1.000000 —- | | |
| LTO | 0.289699 (0.0106) | 0.322902 (0.0042) | 0.413263 (0.0002) | 0.436319 (0.0001) | 0.051276 (0.6579) | 0.108150 (0.3491) | 0.064761 (0.5758) | 0.081397 (0.4816) | 1.000000 —- | |
| IVR | 0.330275 (0.0034) | 0.295844 (0.0090) | 0.240250 (0.0353) | 0.226169 (0.0479) | -0.375957 (0.0008) | 0.296610 (0.0088) | −0.004143 (0.9715) | −0.182978 (0.1112) | −0.450619 (0.0000) | 1.000000 —- |

Note: *p*-value in parentheses. Source: composed by authors.

In addition, as illustrated in Table A1, from variance inflation factors (VIF), which are all lower than 2.5 (a highly strict threshold to verify multicollinearity), we can say that our model does not have an issue of multicollinearity (Johnston et al. 2018). The equation of our econometric model is as follows:

$$Ln\ GII_{07(09)\ i} = \beta_0 + \beta_1 PWI_i + \beta_2 IDV_i + \beta_3 MAS_i + \beta_4 UAI_i + \beta_5 LTO_i + \beta_6 IVR_i + \varepsilon_i \quad (1)$$

$$GII_{19(21)} = \beta_0 + \beta_1 PWI_i + \beta_2 IDV_i + \beta_3 MAS_i + \beta_4 UAI_i + \beta_5 LTO_i + \beta_6 IVR_i + \varepsilon_i \quad (2)$$

where Ln_07(09) and GII_19(21) are dependent variables and denote (natural logarithm of) Global Innovation Index for a year (07, 09) 19 and 21. Our study selected 2007 and 2009 as the pre-crisis period and 2019 and 2021 as the post-crisis period (for the financial crisis and COVID-19 pandemic, respectively). The six indices of the HCD are used as explanatory variables in our study. PWI denotes power distance index (a scale of 0–100), and 0 indicates low power distance, while 100 indicates high power distance. IDV denotes a level of individualism (a scale of 0–100). A score close to 0 indicates collectivism, while that close to 100 indicates individualism. MAS denotes a level of masculinity (a scale of 0–100). A score close to 0 means a feministic society, while that close to 100 means a masculine society. UAI denotes uncertainty avoidance index (a scale of 0–100), and 0 indicates a society of a low level of uncertainty avoidance, while 100 indicates a society of a high level of uncertainty avoidance. LTO denotes long-term orientation (a scale of 0–100), and 0 indicates a (short-term) normative society, while 100 indicates a (long-term) pragmatic society. IVR denotes a level of indulgence (a scale of 0–100). A culture whose score is close to 0 restrains free gratification of natural human needs but a score of to 100 allows that. $\beta_0$ is a constant and $\varepsilon$ is an error term.

In terms of econometric methodology, this study adopts OLS and robust estimations for a multiple regression analysis. In terms of values in skewness and kurtosis, our variables are ranged to be accepted as a normal distribution, although IDV rejects the null hypothesis of the J-B test. The VIF confirms the non-issue of multicollinearity. Above all, our models do not present heteroskedasticity under the OLS regression analysis at a 5% significance level (excluding GII_19 OLS model for developing countries). We tried weighted least square (WLS), a method, which is typically adopted to mitigate the heteroskedasticity, regression analysis by weighting IDV, but contrary to a former expectation (which improves the p-value of Breusch–Pagan test) the adoption of WLS caused the heteroskedasticity, which has not manifested under the OLS model. Thereby, we confirmed that OLS better performs in our models. However, to confirm the robustness of the regression outputs, we further additionally constructed a model using robust least square (M-estimation), which is less sensitive to outliers and normality, by employing the least median of squares method (Massart et al. 1986; Rousseeuw 1990).

## 4. Results and Discussion

To clarify the general relation of cultural dimensions and innovation development and distinctiveness between developed and developing income groups, we ran regression analyses for the three types of models: all countries (77), developed countries (39) and developing countries (38) (See Table A2). In addition, an inclusion of four different years of the GII scores into the models enables us to test changes in these relations in pre- and post-crisis periods and throughout the oldest and latest datasets of GII scores in 2007 and 2021.

As illustrated in Table 5, the regression results for all countries are consistent regardless of time variations and regression methodologies, which shows a strong association of the HCD with the GII scores with statistical significance (except for MAS). This indicates that, in general, the same cultural properties have been required for countries to enhance innovation inputs and outputs from 2007 to 2021. Unlike our hypothesis, countries, which achieved high innovation development are defined by the same cultural properties somehow asserted in the previous studies, namely, low power distance (Shane 1993; Rinne et al. 2012; Prim et al. 2017; Espig et al. 2021), individualism (Shane 1993; Rinne et al. 2012; Taylor and Wilson 2012; Cox and Khan 2017; Espig et al. 2021), low uncertainty avoidance (Shane 1993; Prim et al. 2017; Espig et al. 2021), long-term orientation (Cox and Khan 2017; Prim et al. 2017; Espig et al. 2021) and high indulgency (Prim et al. 2017; Espig et al. 2021) regardless of pre- and post-crisis period and time variances (rejection of H1, H2 and H3).

**Table 5.** A cross-sectional regression output (all countries).

| Dependent Variables | Ln(GII_07) | | Ln(GII_09) | | GII_19 | | GII_21 | |
|---|---|---|---|---|---|---|---|---|
| Models | OLS | Robust | OLS | Robust | OLS | Robust | OLS | Robust |
| PWI | −0.002854 ** | −0.002906 * | −0.003060 *** | −0.002865 *** | −0.162910 *** | −0.155237 *** | −0.169453 *** | −0.159133 *** |
| | (0.001413) | (0.001485) | (0.000995) | (0.000948) | (0.053950) | (0.056031) | (0.057668) | (0.060184) |
| IDV | 0.004258 *** | 0.003909 *** | 0.002960 *** | 0.003718 *** | 0.165820 *** | 0.176261 *** | 0.156257 *** | 0.167569 *** |
| | (0.001250) | (0.001314) | (0.000880) | (0.000838) | (0.047733) | (0.049575) | (0.051023) | (0.053249) |
| MAS | 0.000856 | 0.000730 | −0.000742 | −0.000989 | −0.025795 | −0.027511 | −0.029463 | −0.032742 |
| | (0.001030) | (0.001083) | (0.000725) | (0.000691) | (0.039338) | (0.040856) | (0.042049) | (0.043884) |
| UAI | −0.002684 *** | −0.002552 *** | −0.001778 *** | −0.000906 | −0.100386 *** | −0.096704 *** | −0.093758 ** | −0.093036 ** |
| | (0.000893) | (0.000939) | (0.000629) | (0.000599) | (0.034115) | (0.035431) | (0.036466) | (0.038057) |
| LTO | 0.004852 *** | 0.004950 *** | 0.003926 *** | 0.003775 *** | 0.265647 *** | 0.266500 *** | 0.290284 *** | 0.290233 *** |
| | (0.000958) | (0.001008) | (0.000675) | (0.000643) | (0.036597) | (0.038009) | (0.039119) | (0.040825) |
| IVR | 0.003623 *** | 0.003847 *** | 0.002337 *** | 0.002539 *** | 0.127366 *** | 0.135438 *** | 0.137386 *** | 0.144106 *** |
| | (0.001066) | (0.001121) | (0.000750) | (0.000715) | (0.040705) | (0.042275) | (0.043510) | (0.045408) |
| Constant | 0.791623 *** | 0.786725 *** | 1.154604 *** | 1.051247 *** | 34.25055 *** | 32.68135 *** | 31.67457 *** | 30.26788 *** |
| | (0.157226) | (0.165321) | (0.110713) | (0.105468) | (6.004710) | (6.236364) | (6.418554) | (6.698551) |
| Breusch-Pagan *p*-value | 0.7766 | - | 0.0719 | - | 0.3254 | - | 0.3097 | - |
| $R^2$ adjust. | 0.647129 | 0.566746 | 0.694742 | 0.632580 | 0.720600 | 0.658437 | 0.701010 | 0.635567 |
| Obs. | 77 | 77 | 77 | 77 | 77 | 77 | 77 | 77 |

Note: Standard errors in parentheses; * *p*-value < 0.1; ** *p*-value < 0.05; *** *p*-value < 0.01. Source: composed by authors.

This result points out several notable elaborations on the possible connection of cultural dimensions with national innovation output. Firstly, it is interesting to note that despite the time variation and longevity of analysis period, the correlations and cultural dimension values remained the same, and even the crisis events such as 2009, 2014 and 2020 could not change this pattern. This can be attributed to the strong mainstream beliefs in governments of many countries. Some (mostly developed) states and their corporations believe that innovations are the key to overcoming crises and thus stabilize or even increase appropriate spending in troubled times (Okoń-Horodyńska 2021). Although the culture properties outlined in the analysis above generally are associated with "innovative behavior", there is a hurdle to follow this path in our analysis. The issue is that among the top 15 GII's most innovative states, there are several countries with cultural properties that almost completely mirror the above mentioned set, namely South Korea, Singapore, Japan and China. The high GII values of these countries do not change the complete statistical picture of the analysis but provide an invaluable proof that the culture itself does not explicitly support high innovation activity.

On the other hand, contrary to some previous studies, which demonstrated a negative association of MAS with the innovation development (Cox and Khan 2017; Prim et al. 2017; Espig et al. 2021) in our models, MAS consistently shows no statistical significance. Interestingly, its coefficient signs even show positive in the year 2007 and 2009, both in OLS and Robust models.

We further examined the case by dividing the income group into the developed and developing to see the impacts of the importance of culture on the GII scores in different periods. The results from the models composed of the 39 developed countries present both similarity and difference compared to that for all the 77 countries (See Table 6). The results are consistent in terms of a time variance, which implies that both financial and pandemic crisis do not influence the relationship between the HCD and the GII scores, and the passage of the time from 2007 to 2021 also does not mean that (rejection of H1, H2 and H3).

**Table 6.** A cross-sectional regression output (developed countries).

| Dependent Variables | Ln(GII_07) | | Ln(GII_09) | | GII_19 | | GII_21 | |
|---|---|---|---|---|---|---|---|---|
| Models | OLS | Robust | OLS | Robust | OLS | Robust | OLS | Robust |
| PWI | −0.000778 | −0.001409 | −0.001446 | −0.001347 | −0.076265 | −0.072946 | −0.075039 | −0.070211 |
|  | (0.001791) | (0.001774) | (0.001036) | (0.001070) | (0.062185) | (0.069352) | (0.068265) | (0.076067) |
| IDV | 0.001164 | 0.000263 | −0.000560 | −0.000315 | 0.026027 | 0.025094 | 0.015890 | 0.018681 |
|  | (0.001516) | (0.001502) | (0.000877) | (0.000906) | (0.052634) | (0.058700) | (0.057780) | (0.064383) |
| MAS | 0.001629 | 0.001464 | $6.06 \times 10^{-5}$ | −0.000110 | 0.004078 | 0.003995 | 0.000295 | −0.001310 |
|  | (0.001059) | (0.001049) | (0.000613) | (0.000633) | (0.036764) | (0.041002) | (0.040359) | (0.044972) |
| UAI | −0.003347 *** | −0.003236 *** | −0.002832 *** | −0.002438 *** | −0.150495 *** | −0.151523 *** | −0.144954 *** | −0.151368 *** |
|  | (0.001110) | (0.001099) | (0.000642) | (0.000663) | (0.038537) | (0.042978) | (0.042305) | (0.047140) |
| LTO | 0.003923 *** | 0.004378 *** | 0.002446 *** | 0.002573 *** | 0.189846 *** | 0.203242 *** | 0.216101 *** | 0.233468 *** |
|  | (0.001373) | (0.001359) | (0.000794) | (0.000820) | (0.047651) | (0.053143) | (0.052310) | (0.058289) |
| IVR | 0.005177 *** | 0.005137 *** | 0.003972 *** | 0.004396 *** | 0.196409 *** | 0.203497 *** | 0.203945 *** | 0.204453 *** |
|  | (0.001691) | (0.0016740 | (0.000978) | (0.001010) | (0.058690) | (0.065454) | (0.064428) | (0.071791) |
| Constant | 0.890554 *** | 0.937889 *** | 1.356864 *** | 1.289585 *** | 42.75596 *** | 41.60606 *** | 39.93662 *** | 38.90452 *** |
|  | (0.219765) | (0.217644) | (0.127121) | (0.131303) | (7.629076) | (8.508373) | (8.375022) | (9.332166) |
| Breusch-Pagan *p*-value | 0.5221 | — | 0.2439 | — | 0.7826 | — | 0.8559 | — |
| $R^2$ adjust. | 0.510528 | 0.455223 | 0.645048 | 0.605178 | 0.621160 | 0.512370 | 0.569588 | 0.471913 |
| Obs. | 39 | 39 | 39 | 39 | 39 | 39 | 39 | 39 |

Note: Standard errors in parentheses; *** *p*-value < 0.01. Source: composed by authors.

On the other hand, it is worth noting that PWI and IDV lost their statistical significance if making a comparison only between the developed countries. This indicates that generally low power distance and individualism are determining factors to promote the innovation development but only up to a certain level. These two variables are rather more important to sustain its level of innovation development compared to other significant cultural variables, namely UAI, LTO and IVR, and it also means that not the crisis itself but the income group is the factor that influences the relationship of power distance and individualism with the GII scores. To elaborate this point, the aversion of uncertainty, pragmatism and unrestrainedness show a strong positive coefficient with statistical significance, which implies that these are much determining factors to be outstanding in the GII scores amongst the developed income group relative to a low power distance and individualism.

This can be a slightly more predictable result, as developed countries in the GII list generally share the low power distance and individualism factors, thus showing a higher innovation result statistically relies on other cultural variables. However, the important outcome is that developed countries generally rely on IDV and PWI properties to sustain their position, which implies that these properties bear some attributed social benefits that provide a generic support for innovation systems in the country.

Additionally, from the results of models composed of the 38 developing countries, it also turned out that the crisis does not have any influence on the relationship between the HCD and the GII scores as the patterns between 2007 and 2009 and those between 2019 and 2021 are similar (See Table 7), which is in line with those factors for all and for developed countries.

**Table 7.** A cross-sectional regression output (developing countries).

| Dependent Variables | Ln(GII_07) | | Ln(GII_09) | | GII_19 | | GII_21 | |
|---|---|---|---|---|---|---|---|---|
| Models | OLS | Robust | OLS | Robust | OLS | Robust | OLS | Robust |
| PWI | −0.002035 | −0.002142 | 0.000260 | −0.001009 | −0.033735 | −0.015162 | −0.064667 | -0.031370 |
|  | (0.002511) | (0.002678) | (0.001253) | (0.000816) | (0.085837) | (0.088294) | (0.096758) | (0.100343) |
| IDV | 0.005508 ** | 0.005182 ** | 0.003016 ** | 0.002523 *** | 0.126414 | 0.146809 * | 0.130779 | 0.166497 * |
|  | (0.002345) | (0.002501) | (0.001170) | (0.000762) | (0.080152) | (0.082446) | (0.090350) | (0.093698) |
| MAS | −0.000732 | −0.000764 | −0.000101 | 0.001085 | 0.037040 | 0.021645 | 0.019531 | −0.015232 |
|  | (0.002518) | (0.002686) | (0.001257) | (0.000819) | (0.086081) | (0.088545) | (0.097033) | (0.100629) |
| UAI | −0.003249 ** | −0.003206 ** | −0.001198 * | −0.000596 | −0.067213 | −0.044761 | −0.066495 | −0.026960 |
|  | (0.001382) | (0.001475) | (0.000690) | (0.000450) | (0.047260) | (0.048613) | (0.053273) | (0.055248) |
| LTO | 0.003696 ** | 0.003721 ** | 0.002151 *** | 0.003709 *** | 0.198665 *** | 0.173618 *** | 0.229556 *** | 0.181481 *** |
|  | (0.001480) | (0.001579) | (0.000739) | (0.000481) | (0.050611) | (0.052060) | (0.057051) | (0.059165) |
| IVR | 0.002798 ** | 0.002939 ** | 0.000827 | 0.001197 *** | 0.061644 | 0.055528 | 0.078448 | 0.068854 |
|  | (0.001318) | (0.001405) | (0.000657) | (0.000428) | (0.045042) | (0.046331) | (0.050772) | (0.052654) |
| Constant | 0.854685 *** | 0.862729 *** | 0.919344 *** | 0.808375 *** | 23.52133 ** | 21.85161 ** | 22.97308 ** | 20.48578 * |
|  | (0.262217) | (0.279697) | (0.130854) | (0.085260) | (8.964136) | (9.220766) | (10.10468) | (10.47913) |
| Breusch-Pagan *p*-value | 0.1099 | - | 0.1621 | - | 0.0424 | - | 0.0514 | - |
| $R^2$ adjust. | 0.267857 | 0.211523 | 0.285496 | 0.266592 | 0.319472 | 0.194222 | 0.302659 | 0.157053 |
| Obs. | 38 | 38 | 38 | 38 | 38 | 38 | 38 | 38 |

Note: Standard errors in parentheses; * *p*-value < 0.1; ** *p*-value < 0.05; *** *p*-value < 0.01. Source: composed by authors.

However, it presents much dynamics in terms of time variations. In the line with the above results for all countries and developed countries, pragmatism is a strong determinant of the GII scores for the developing countries in all periods, while the importance of IDV, UAI and IVR fades out (either) in OLS and (or) Robust regression analyses for the GII scores in 2019 and 2021, which indicates that the cultural dimensions have recently become less of an influence on the GII scores (Support of H4). This can be the additional evidence of the dominance of non-cultural factors in developing countries' innovation practices as measured by GII.

## 5. Conclusions

This study investigates the impact of the Hofstede cultural dimensions (HCD) on the Global Innovation Index (GII) scores in four different years (2007, 2009, 2019 and 2021) to compare the impacts during pre- and post-crisis (financial and COVID-19) periods. Our study showed several interesting patterns that can be interpreted in several ways. We can see that most of our hypotheses were rejected by the results, except for H4. The more generic conclusions point out that culture, although being an important component of innovative output of a country, does not have a serious impact on the GII results of developing countries. It can be elaborated that properties of IDV and PDI are often associated with countries that have access to slightly more financial, human and educational resources and thus the price of a mistake (meaning "spending on fruitless innovations") is lower than in poorer regions. Therefore, the social and governmental structure there provide fewer barriers in the way of innovators, thus, in the long term, leading to higher innovation output measured by GII. However, this hypothesis requires additional research and analysis.

The far more straightforward explanation of the analysis outcome is that modern innovation systems require a large amount of infrastructural, legal, educational and financial investment from the government. Once these investments are made, the impact of culture properties becomes secondary to the availability of venture and governmental funding, educational and laboratory facilities, copyright protection, etc. Therefore, while cultural properties can theoretically impact the number of innovators in the constructed innovation systems, the systemic output measured by GII would not be drastically impacted by them. Theoretically, this implies that desirable cultural properties for innovation development are the same regardless of crisis.

Finally, the more specific analysis outcome is the absence of crisis impact on cultural impact in the study. This can clearly be attributed to the fact of stable or even increased funding of the appropriate innovation systems by developed countries over the last 15 years. It is also important to remember that since 2009, crises the USA, Japan and EU states have launched extensive "quantitative easing" programs that allowed the flooding of the economies with low-cost credit and therefore provided monetary support for venture capital funding innovation systems. At the same time, while the developed countries continue to support their prolonged investments in their innovation systems, the developing countries could not allocate such resources for this cause. Thus, the results in GII of most developing states remained more or less low, with even less funding allocation during the COVID-19 pandemic. It is highly likely that in terms of innovation, no cultural development or change can significantly impact the innovation output of developing countries without the construction of the appropriate systems. Theoretically, the results confirmed a significance of the income level on innovation development. Practically, the results imply that the developing countries' governments should develop a sophisticated innovation system to a certain level as the first priority for innovation development.

On the other hand, we acknowledge that this study has potential limitations. The number of countries included in the model is rather small due to the constrained datasets of Hofstede's cultural dimensions and GII overtime, which can cause a sample bias. Future studies on the impact of cultural factors on innovation development can be made based on expanded sample countries. In addition, we used the total value of GII to evaluate the overall impact of culture on innovation. However, GII is composed of a complex of various sub-indicators. Thus, the follow-up study can be sophisticated if focusing on impacts on a specific sub-indicator of GII.

**Author Contributions:** Data curation, H.-S.L.; Formal analysis, S.U.C. and S.N.; Investigation, H.-S.L., S.U.C., S.N. and E.A.D.; Methodology, H.-S.L.; Software, H.-S.L.; Validation, S.U.C. and E.A.D.; Visualization, S.N. All authors have read and agreed to the published version of the manuscript.

**Funding:** The article was prepared with the financial support of the RFFR as part of a research project 'Opportunities and prospects for the development of strategic alliances of innovative organizations in Hungary and Russia in the field of biotechnology and pharmaceuticals', project N° 21-510-23004.

**Institutional Review Board Statement:** Not applicable.

**Informed Consent Statement:** Not applicable.

**Data Availability Statement:** Not applicable.

**Conflicts of Interest:** The authors declare no conflict of interest.

## Appendix A

**Table A1.** A variance inflation factor (VIF) test.

|     | All Countries | Developed Countries | Developing Countries |
|-----|---------------|---------------------|----------------------|
| PWI | 2.438042      | 2.149355            | 1.392218             |
| IDV | 2.370354      | 1.791110            | 1.142869             |
| MAS | 1.080714      | 1.177884            | 1.166662             |
| UAI | 1.089466      | 1.248804            | 1.113357             |
| LTO | 1.378003      | 1.433472            | 1.794596             |
| IVR | 1.536238      | 1.886875            | 1.457669             |

Source: composed by authors.

Table A2. Countries included in this study.

| Developed Countries (39 Countries) | Developing Countries (38 Countries) |
|---|---|
| Australia, Austria, Belgium, Canada, Chile, Croatia, Czech Republic, Denmark, Estonia, Finland, France, Germany, Greece, Hong Kong, Hungary, Iceland, Ireland, Italy, Japan, Latvia, Lithuania, Luxembourg, Malta, Netherlands New Zealand, Norway, Poland, Portugal, Romania, Singapore, Slovakia, Slovenia, South Korea, Spain, Sweden, Switzerland, UK, USA, Uruguay | Albania, Argentina, Armenia, Azerbaijan, Bosnia and Herzegovina, Brazil, Bulgaria, China, Colombia, Dominican Republic, Georgia, Indonesia, Jordan, Kazakhstan, Malaysia, Mexico, North Macedonia, Paraguay, Peru, Russia, South Africa, Thailand, Turkey, Algeria, Bangladesh, Bolivia, Burkina Faso, Egypt, El Salvador, India, Morocco, Mozambique, Nigeria, Pakistan, Philippines, Tanzania, Ukraine, Vietnam |

Note: The classification of the countries' income group is in accordance with the GII 2021 reports p. 25 (developed countries—high-income group; developing countries—upper-middle income, low–middle income and low-income groups). Countries excluded in this study due to data insufficiency (although they are listed in the HCD) are as follows: Bhutan, Costa Rica, Ecuador, Ethiopia, Fiji, Guatemala, Honduras, Israel, Jamaica, Kenya, Kuwait, Malawi, Namibia, Nepal, Panama, Qatar, Senegal, Sierra Leone, Sri Lanka, Suriname, Syria, Tunisia, UAE. Source: composed by authors.